# The OSU Scheme for Congestion Avoidance in ATM Networks Using Explicit Rate Indication[1]

Raj Jain, Shiv Kalyanaraman and Ram Viswanathan[2]

Department of Computer and Information Science

The Ohio State University

Columbus, OH 43210-1277

Email: {*jain, shivkuma*}*@cis.ohio-state.edu, ramv@microsoft.com*

WWW: *http://www.cis.ohio-state.edu/˜jain*

**Abstract**

We propose an end-to-end rate-based congestion avoidance scheme for ABR traffic on ATM networks using explicit rate indication to sources. The scheme uses a new congestion detection technique and an O(1) switch algorithm to provide high thoughput, low queues, fair operation, quick convergence and a small set of well understood parameters.

## 1 Introduction

Congestion occurs in computer networks whenever the total input traffic is greater than the capacity. When congestion occurs, the sources learn about this condition after a time delay and reduce their traffic. Data may be lost during this time delay for the sources to respond. The amount of data that can be lost depends on the delay-bandwidth product of the link. The traditional goals of congestion control schemes were to achieve low loss, high throughput and low delay.

[2]Ram Viswanathan is currently with Microsoft Corp., WA

High speed networks have higher delay-bandwidth products than low speed networks. In particular, the bandwidth has increased, but the propogation delay has remained the same. Hence, more data can be sent (and lost) before the sources learn about congestion. The problem of congestion is therefore more important in high speed networks (HSNs) particularly ATM networks [3, 11, 12]

Until recently, congestion control has been based on window flow controls (eg., in TCP/IP). Feedback was either implicit (e.g., via timeouts in TCP/IP [4]) or explicit but binary (e.g., in DECbit [7] or its derivatives). Even the early work on ATM congestion control used a explicit but binary feedback method called "Explicit Forward Congestion Indication (EFCI)"[5].

However, an advantage of increased bandwidth is that more control information can be sent at the same percentage overhead. Further, HSNs are connection oriented and the network can maintain state about every connection and can calculate the exact rate the sources should send at. The ATM Forum has therefore decided to adopt a "explicit rate" approach, where the switches can specify the exact rate for the source in its feedback, instead of a single bit. With accurate feedback, a scheme can achieve quick convergence and fairness without increasing the complexity of switch design. We present one such scheme in this paper, the Ohio State University (OSU) scheme.

The OSU scheme is named thus because it was a follow on to MIT scheme [1, 2]. This paper provides an overview of the OSU scheme. The scheme is discussed and analysed in greater detail in [9]. In this paper, we first describe the control cell format, the source, switch and destination algorithms. We then highlight some of the unique features of the OSU scheme which have hence become commonly accepted parts of switch schemes or have been adopted by the standard. The simulation section provides a set of simulation results which illustrate the efficiency, fairness and quick convergence of the scheme in LAN and WAN scenarios. Appendix A gives the pseudo-code for the scheme.

## 2 The OSU Scheme

The OSU scheme requires sources to monitor their load and send control cells *periodically* at intervals of $T$ microseconds. These control cells contain source rate information. The switches monitor their own load and use it with the information provided by the control cells to compute a factor by which the source should go up or down. The destination simply returns the control cells to the source, which then adjusts its rate as instructed by the network. This section described the various components of the scheme.

### 2.1 Control Cell Format

The control cell contains the following the fields relevant to our discussion:

1) Transmission Cell Rate (TCR). The TCR is the inverse of the minimum inter-cell transmission time and indicates instantaneous peak load input by the source.

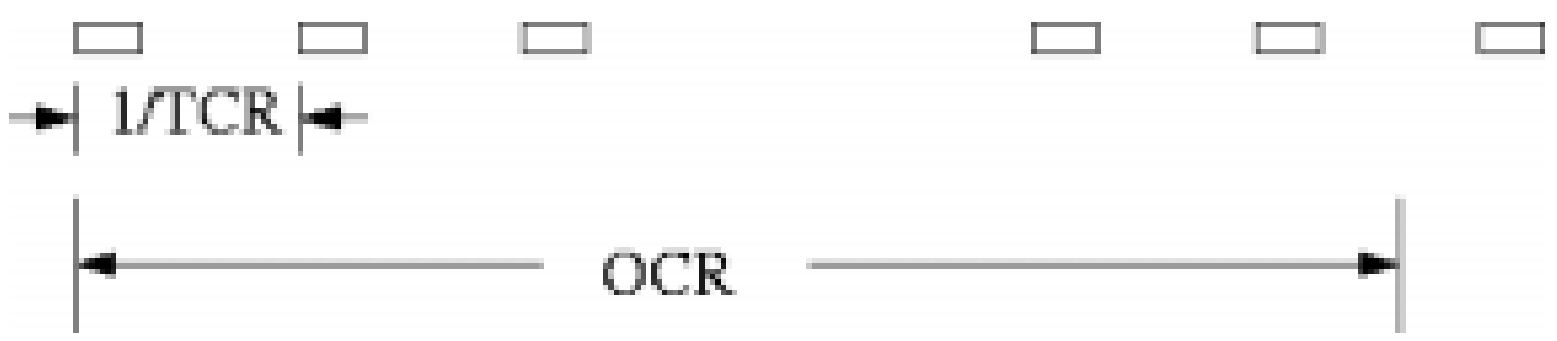

Figure 1: Transmitted cell rate (instantaneous) and Offered Average Cell Rate (average).

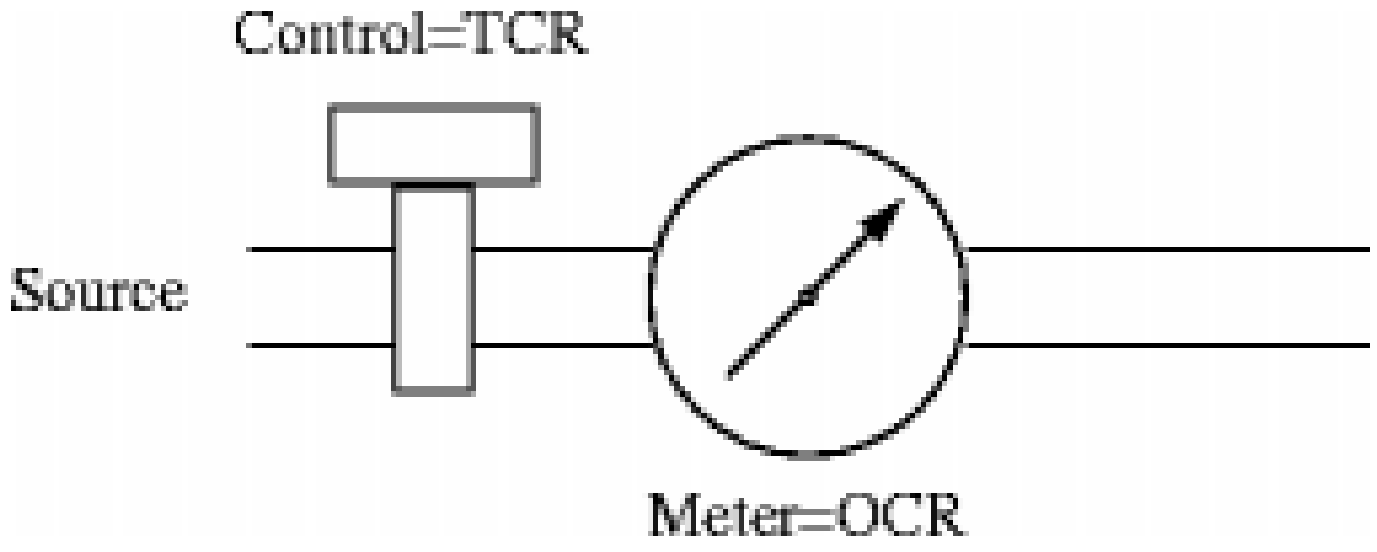

Figure 2: Transmitted cell rate (controlled) and Offered Average Cell Rate (measured).

2) Offered Cell Rate (OCR). For bursty sources which may not send a cell at every transmission opportunity, TCR is not a good indication of overall load. Therefore, the average *measured* load over $T$ interval is indicated in the OCR field of the control cell. The inter-cell time is computed based on the transmitted cell rate. However, the source may be idle in between the bursts and so the average cell rate is different from the transmitted cell rate. This average is called the offered average cell rate and is also included in the cell. This distinction between TCR and OCR is shown in Figure 1. Notice that TCR is a control variable (like the knob on a faucet) while the OCR is a measured quantity (like a meter on a pipe). This analogy is shown in Figure 2. The initialization of TCR and OCR are discussed in Section 2.2.1.

3) Load Adjustment Factor (LAF). This field carries the feedback from the network. At the source, the LAF is initialized to zero. Switches on the path can only increase LAF. Increasing the LAF corresponds to decreasing the allowed source rate. Hence, successive switches only reduce the rate allowed to the source. Thus, the source receives the rate allowed by the bottleneck along the path.

4) Averaging interval (AI). The OSU scheme primarily uses measured quantities instead of parameters for control. These quantities are measured at the source (eg., OCR) and the switch (eg., current load level $z$ discussed in section 2.3.1). The measurements are done over intervals (called "averaging intervals") to smoothen out the variance in these quantities. To ensure coorelation of the measured quantities at the switch and at the source, we require the source averaging intervals to be the maximum of the averaging interval of the switches along the path. This maximum value is returned in the AI field. The AI field is initialized to zero at the source. This is further discussed in Section 2.3.1.

## 2.2 The Source Algorithm

In this section, we discuss the source algorithm which consists of two parts: initializing and sending control cells and reacting to network feedback when these control cells return

### 2.2.1 Control Cell Sending Algorithm

The sources send a control cell into the network every $T$ microseconds. The source initializes all the fields. The network reads only the OCR, LAF and AI fields and modifies only the LAF and AI fields. The TCR field is used by the source to calculate the new TCR as discussed in the next section.

LAF and AI are both initialized to zero as discussed in Section 2.1. The OCR field is initialized to the measured load over the last $T$ microseconds (since the sending of the last control cell). As defined earlier, TCR is the inverse of the inter-cell time while OCR is the measured rate. Consistent with this definition, we require that OCR in cell $\leq$ TCR in cell. However, when TCR has just been reduced, the OCR may have a value between the old TCR and current TCR. Hence, we initialize

$$\text{TCR in Control Cell} \leftarrow \max\{\text{TCR, OCR}\}$$

### 2.2.2 Responding to Network Feedback

The source uses the TCR and the modified LAF and AI fields of the returned control cell to calculate its new rate (TCR) as follows:

$$\text{New TCR} \leftarrow \frac{\text{TCR in Cell}}{\text{LAF in Cell}}$$

$$\text{if (LAF} \geq 1 \textit{ and } \text{New TCR} < \text{TCR) TCR} = \text{New TCR}$$

$$\text{else if (LAF} < 1 \textit{ and } \text{New TCR} > \text{TCR) TCR} = \text{New TCR}$$

When LAF $\geq$ 1, the network is asking the source to decrease its TCR. If *New TCR* is less than the current TCR, the source reduces its TCR to *New TCR*. No adjustments are required otherwise. The other case (LAF $<$ 1) is similar.

The source interval $T$ is set to the maximum of the switch averaging intervals in the path which has been returned in the AI field of the control cell. The method ensures that a switch sees atmost one control cell from every source per switch interval. This point is further explained in Section 2.3.1.

## 2.3 The Switch Algorithm

The switch algorithm consists of two parts: measuring the current load level periodically and calculating the feedback whenever a control cell is received. The feedback calculation consists of an algorithm to achieve efficiency and an algorithm to achieve fairness. The measured value of the current load level is used to decide whether the efficiency or the fairness algorithm is used to calculate feedback.

### 2.3.1 Measuring The Current Load Level $z$

The switch measures its current load level, $z$ , as the ratio of its "input rate" to its "target output rate". The input rate is measured by counting the number of cells *received* by the switch during a fixed averaging interval. The target output rate is set to a fraction (close to 100 %) of the link rate. This fraction, called Target Utilization ($U$ ), allows high utilization and low queues in steady state. The current load level $z$ is used to detect congestion at the switch and determine an overload or underload condition.

$$\text{Target Output Cell Rate} = \frac{\text{Target Utilization (U)} \times \text{Link bandwidth in Mbps}}{\text{Cell size in bits}}$$

$$z = \frac{\text{Number of cells received during the averaging interval}}{\text{Target Output Cell Rate} \times \text{Averaging Interval}}$$

The switches on the path have averaging intervals to measure their current load levels ($z$). These averaging intervals are set locally by network managers. A single value of $z$ is assumed to correspond to *one* OCR value of every source. If two control cells of a source with different OCRs are seen in a single interval (for one value of $z$), the above assumption is violated and conflicting feedbacks may be given to the source. So, when feedback is given to the sources the AI field is set to the maximum of the AI field in the cell and the switch averaging interval:

$$\text{AI in cell} \leftarrow \text{Max(AI in cell, switch averaging interval)}$$

### 2.3.2 Achieving Efficiency

Efficiency is achieved as follows:

$$\text{LAF in cell} \leftarrow \text{Max(LAF in cell, } z\text{)}$$

The idea is that if all sources divide their rates by LAF, the switch will have $z = 1$ in the next cycle. In the presence of other bottlenecks, this algorithm converges to $z = 1$. In fact it reaches a band $1 \pm \Delta$ quickly. This band is identified as an efficient operating region. However, it does not ensure fair allocation of available bandwidth among contending sources. When $z = 1$, sources may have an unfair distribution of rates.

### 2.3.3 Achieving Fairness

Our first goal is to achieve efficient operation. Once the network is operating close to the target utilization, we take steps to achieve fairness. The network manager declares a target utilization band (*TUB*), say, 90±9% or 81% to 99%. When the link utilization is in the TUB, the link is said to be operating efficiently. The TUB is henceforth expressed in the U(1±Δ) format, where $U$ is the target utilization and $\Delta$ is the half-width of the TUB. For example, 90±9% is expressed as $90(1 \pm 0.1)\%$. Equivalently, the TUB is identified when the current load level $z$ lies in the interval $1 \pm \Delta$.

We also need to count the number of active sources for our algorithm. The number of active sources can be counted in the same averaging interval as that of load measurement. One simple method is to mark a bit in the VC table whenever a cell from a VC is seen. The bits are counted at the end of each averaging interval and are cleared at the beginning of each interval. Alternatively a count variable could be incremented when the bit is changed from zero to one. This count variable and the bits are cleared at the end of the interval.

Given the number of active sources, a fair share value is computed as follows:

$$\text{FairShare} = \frac{\text{Target Cell Rate}}{\text{Number of Active Sources}}$$

Underloading sources are sources that are using bandwidth less than the FairShare and overloading sources are those that are using more than the FairShare. To achieve fairness, we treat underloading and overloading sources differently. If the current load level is $z$, the underloading sources are treated as if the load level is $z/(1+\Delta)$ and the overloading sources are treated as if the load level is $z/(1-\Delta)$.

$$\text{If (OCR in cell} < \text{FairShare)} \;\; \text{LAF in cell} \leftarrow \text{Max(LAF in cell, } \frac{z}{(1+\Delta)})\}$$

$$\text{else LAF in cell} \leftarrow \text{Max(LAF in cell, } \frac{z}{(1-\Delta)})\}$$

We prove in [9, 10] that this algorithm guarantees that the system consistently moves towards fair operation. We note that all the switch steps are O(1) w.r.t. the number of VCs.

#### 2.3.4 What Load Level to Use ?

The OCR in the control cell is corelated to $z$ when the control cell *enters* the switch queue. The value of $z$ may change before the control cell leaves the switch queue. The OCR in the cell at the time of leaving the queue is not necessarily co-related with $z$. Hence, the above computations are done and feedback give when the control cell enters the queue.

### 2.4 The Destination Algorithm

The destination simply returns all control cells back to the source.

# 3 Unique Features of the OSU scheme

In this section, we highlight some of the unique features of the OSU scheme which have become commonly accepted parts of many other schemes. This includes applying the concept of congestion avoidance to rate-based algorithms and the use of input rate instead of queue length for congestion detection. The number of parameters is small and their effects are well understood.

## 3.1 Congestion Avoidance

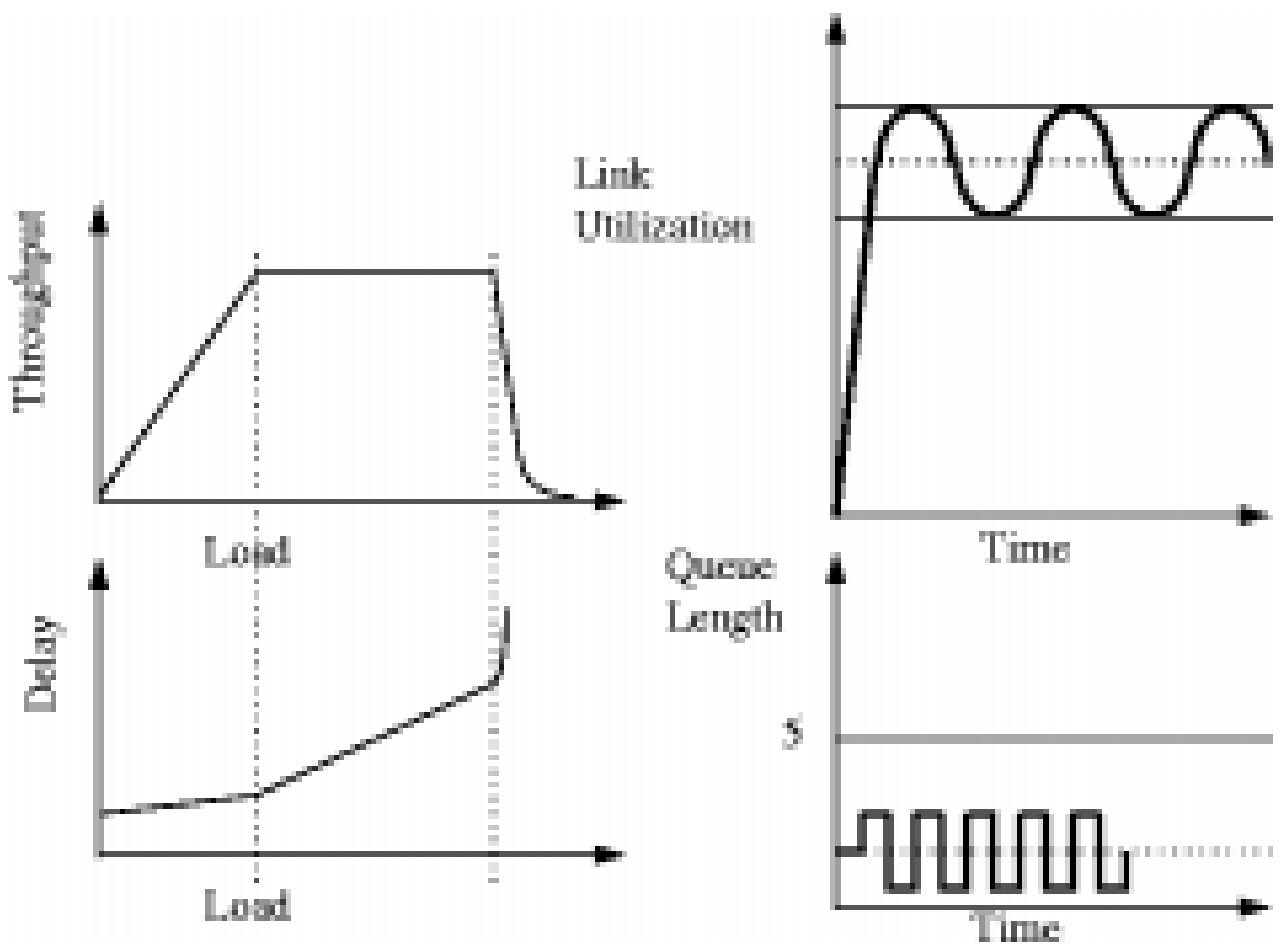

Figure 3: Throughput and delay vs Load

The OSU scheme is a congestion *avoidance* scheme. As defined in [8], a congestion avoidance scheme is one that keeps the network at high throughput and low delay in the steady state. The system operates at the *knee* of the throughput delay-curve as shown in Figure 3.

The OSU scheme keeps the steady state bottleneck link utilization in the target utilization band (TUB). The utilization is high and the oscillations are bounded by the TUB. Hence, in spite of oscillations in the TUB, the load on the switch is always less than one. So the switch queues are close to zero resulting in minimum delay to sources.

## 3.2 Parameters

The OSU scheme requires just three parameters: the switch averaging interval (AI) , the target link utilization ($U$) , and the half-width of the target utilization band ($\Delta$).

The target utilization ($U$) and the TUB present a few tradeoffs. During overload (transients), $U$ affects queue drain rate. Lower $U$ increases drain rate during transients, but reduces utilization in steady state. Further, higher $U$ also constrains the size of the TUB.

A narrow TUB slows down the convergence to fairness (since the formula depends on $\Delta$) but has smaller oscillations in steady state. A wide TUB results in faster progress towards fairness, but has more oscillations in steady state. We find that a TUB of $90\%(1 \pm 0.1)$ used in our simulations is a good choice.

The switch averaging interval affects the stability of $z$. Shorter intervals cause more variation in the $z$ and hence more oscillations. Larger intervals cause slow feedback and hence slow progress towards steady state.

## 3.3 Input Rate vs Queue Length for Congestion Detection

The OSU scheme detects congestion by measuring the current load level based on input rate at the switch queue. Many switch schemes use queue length as the congestion indicator. Queue length is commonly used as the congestion indicator in window-based control. We note that in window-based control, the sum of the source windows equals the maximum queue length. However, in rate-based control, the sum of the source rates (input rate) may be greater than, equal to, or less than the link output rate for *any* value of queue length. Hence, queue length gives no information about the relation between the current input rate and the ideal rate. Rate-based vs window-based control is further discussed in [6].

In rate-based control, the ratio of the input and output rates should be less than one for the switch queues to decrease. Our measure $z$ uses this ratio. We aim for $z = 1$ which guarantees that in steady state, the input rate is smaller than the output rate.

# 4 Simulation Results

We use simple configurations which test efficiency, fairness and speed of convergence. We use the rates of sources, the bottleneck queue lengths and bottleneck link utilization as metrics for our performance evaluation. The distribution of rates shows the fairness between sources. The bottleneck link utilization and queue length complement each other to show the efficiency achieved. The queue length is used as a metric when the link utilization is 100% and the utilization is used as a metric when the queue length is zero. We present only LAN simulations here. A complete set of simulations may be found in [9].

All links are 1 km long running at 155 Mbps. The averaging interval of 300 $\mu$s and a target utilization band of $90(1\pm 0.1)\%$ are used.

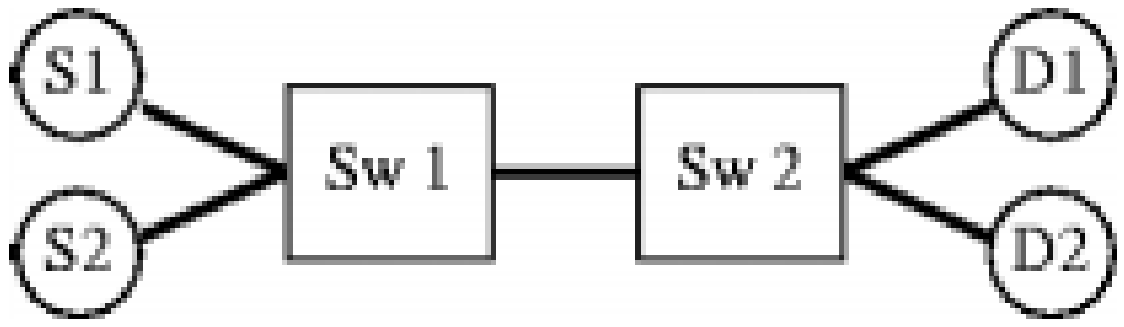

Figure 4: Transient source configuration

## 4.1 Transient Source Simulation

The transient source simulation shown in Figure 4 consists of one persistant connection which is always active. A second persistant connection which shares one inter-switch link with the first, comes on after one third of the simulation run and goes off at two third of the total simulation time.

Initially, there is no bottleneck and the source reaches the maximum rate quickly. When the second source comes on, the two sources quickly converge to their fairshares (half the maximum rate). When the transient source goes away, the first source quickly comes back to maximum rate. This sample configuration tests the steady state as well as the transient response of the scheme. It also shows convergence to fairshares. The TCRs of sources, bottleneck queue length and link utilization are shown in Figure 5.

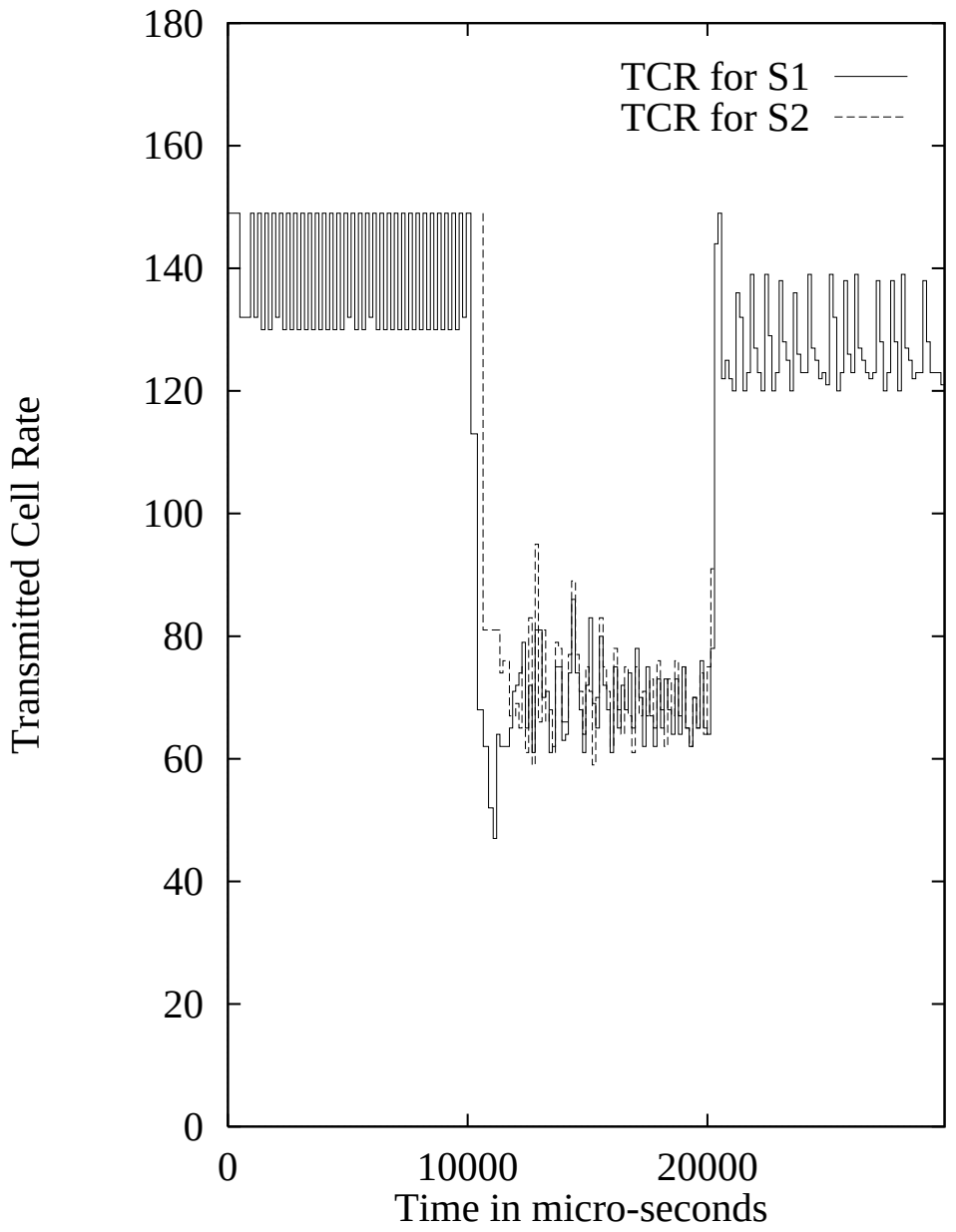

(a) Transmitted Cell Rates

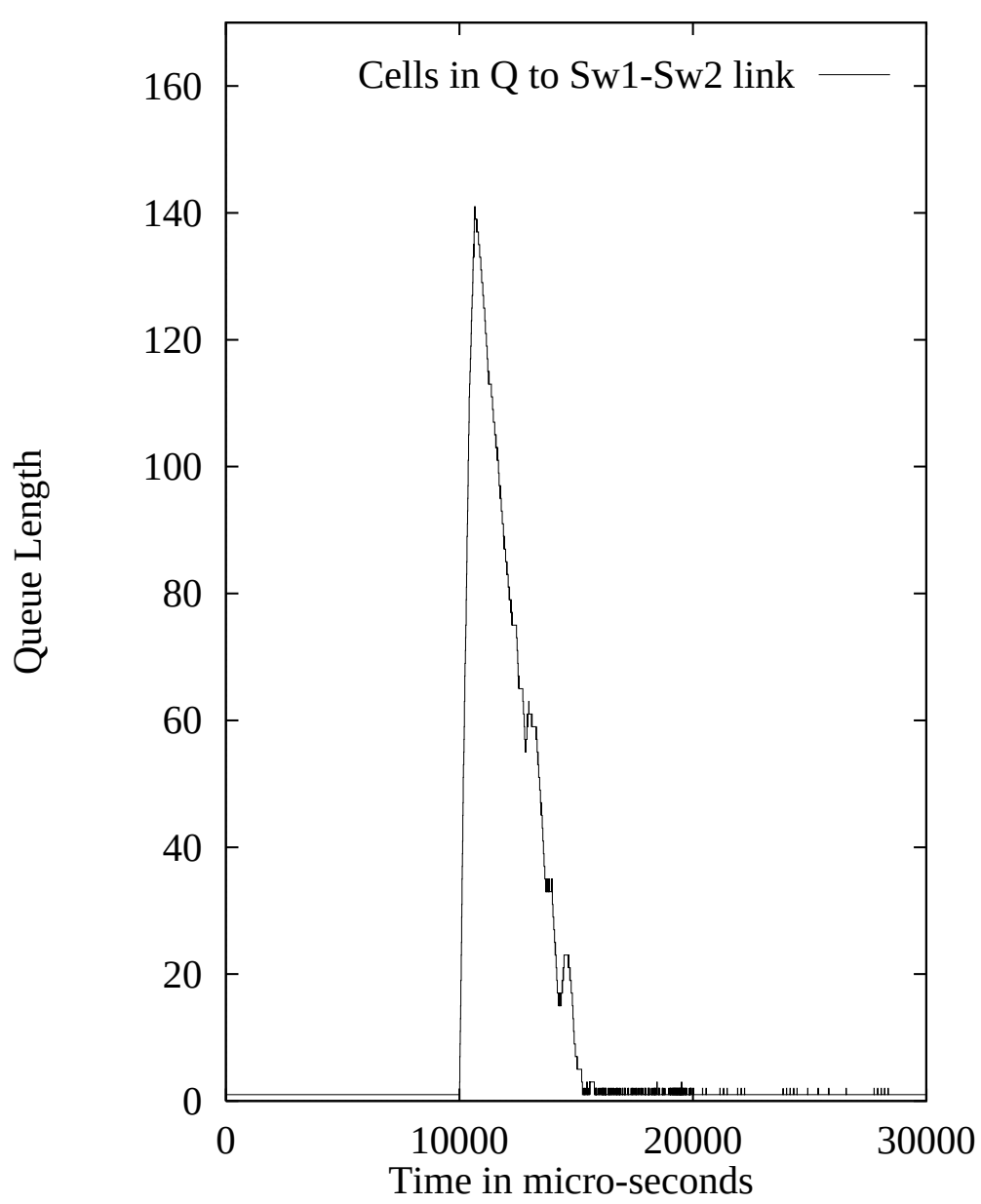

(b) Queue Lengths

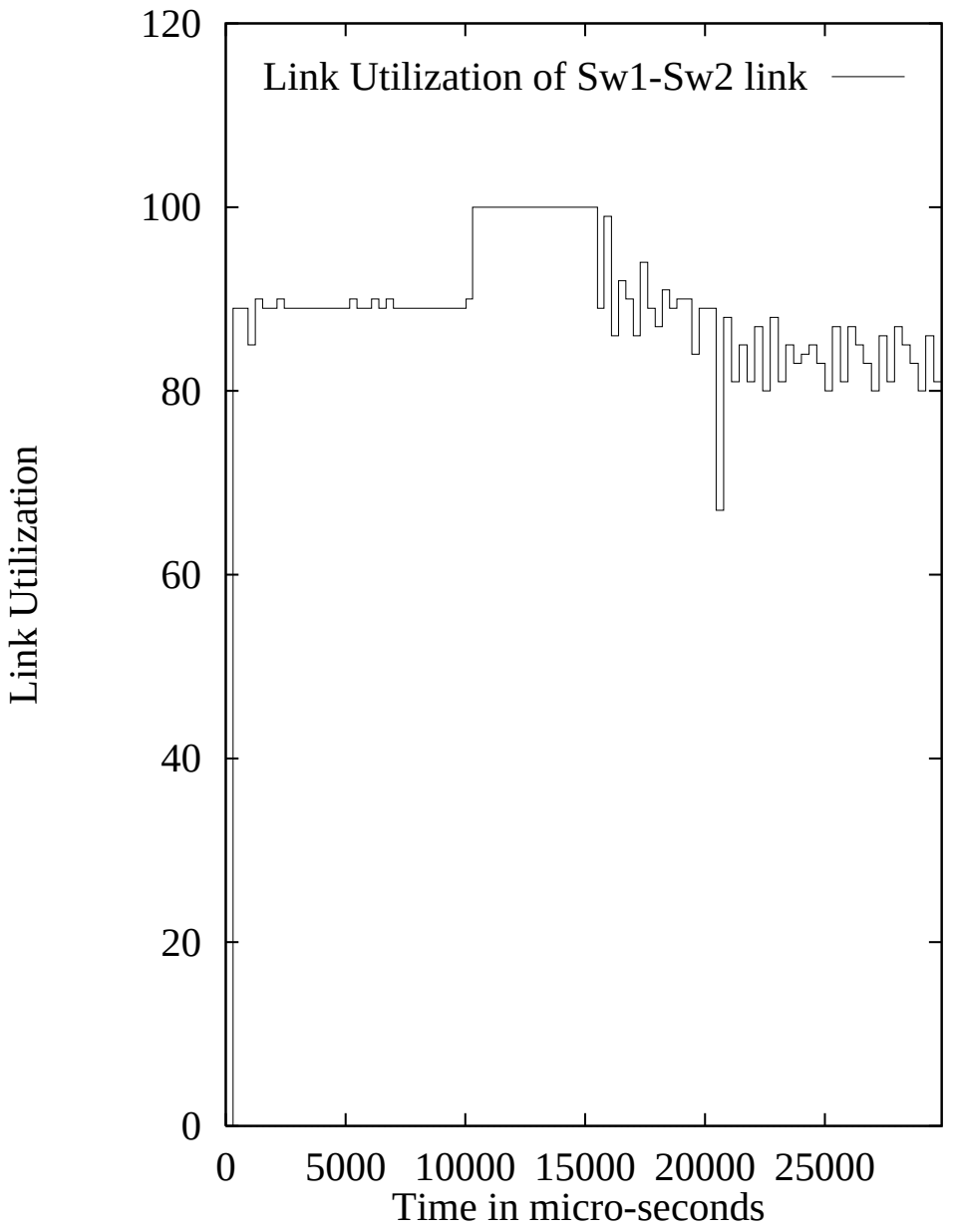

(c) Link Utilization

Figure 5: Simulation results for the Transient Source Simulation

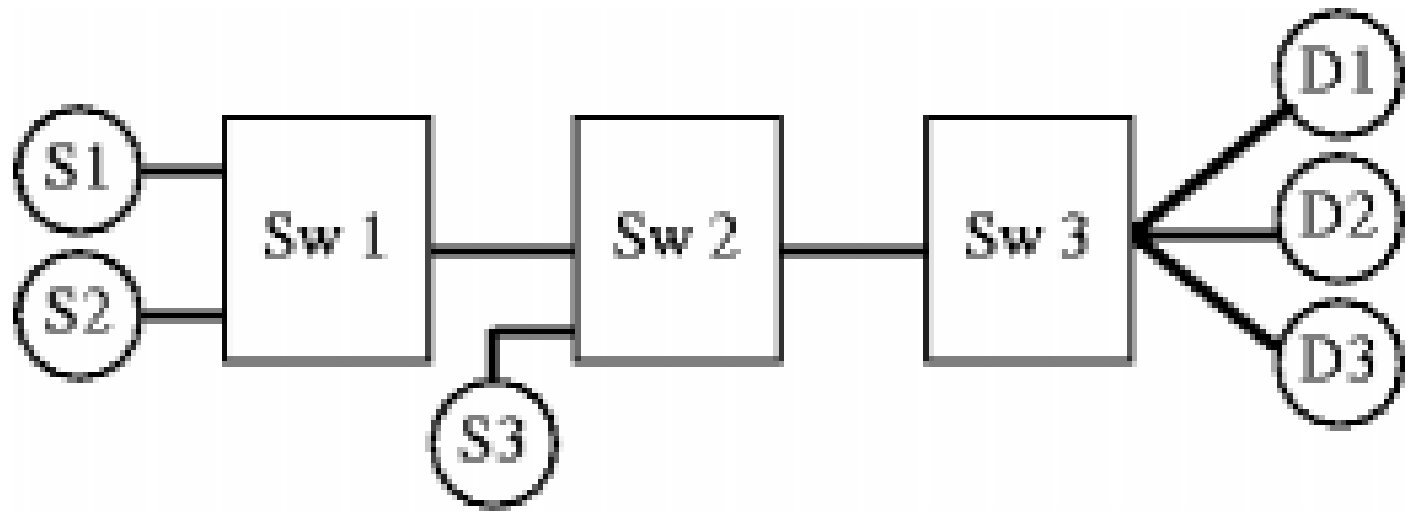

Figure 6: The parking lot configuration

## 4.2 Parking Lot

This configuration is popular for studying fairness. The configuration and its name was derived from theatre parking lots, which consist of several parking areas connected via a single exit path. At the end of the show, congestion occurs as cars exiting from each parking area try to join the main exit stream.

For computer networks, an $n$-stage parking lot configuration consists of $n$ switches connected in a series. There are $n$ VCs. The first VC starts from the first switch and goes to the end. For the remaining $i$th VC starts at the $i-1$th switch. All users should get the same throughput regardless of the parking area used. A 3-switch parking lot configuration is shown in Figure 6. The simulation results are shown in Figure 7. Notice that all VCs receive the same throughput without any fair queueing.

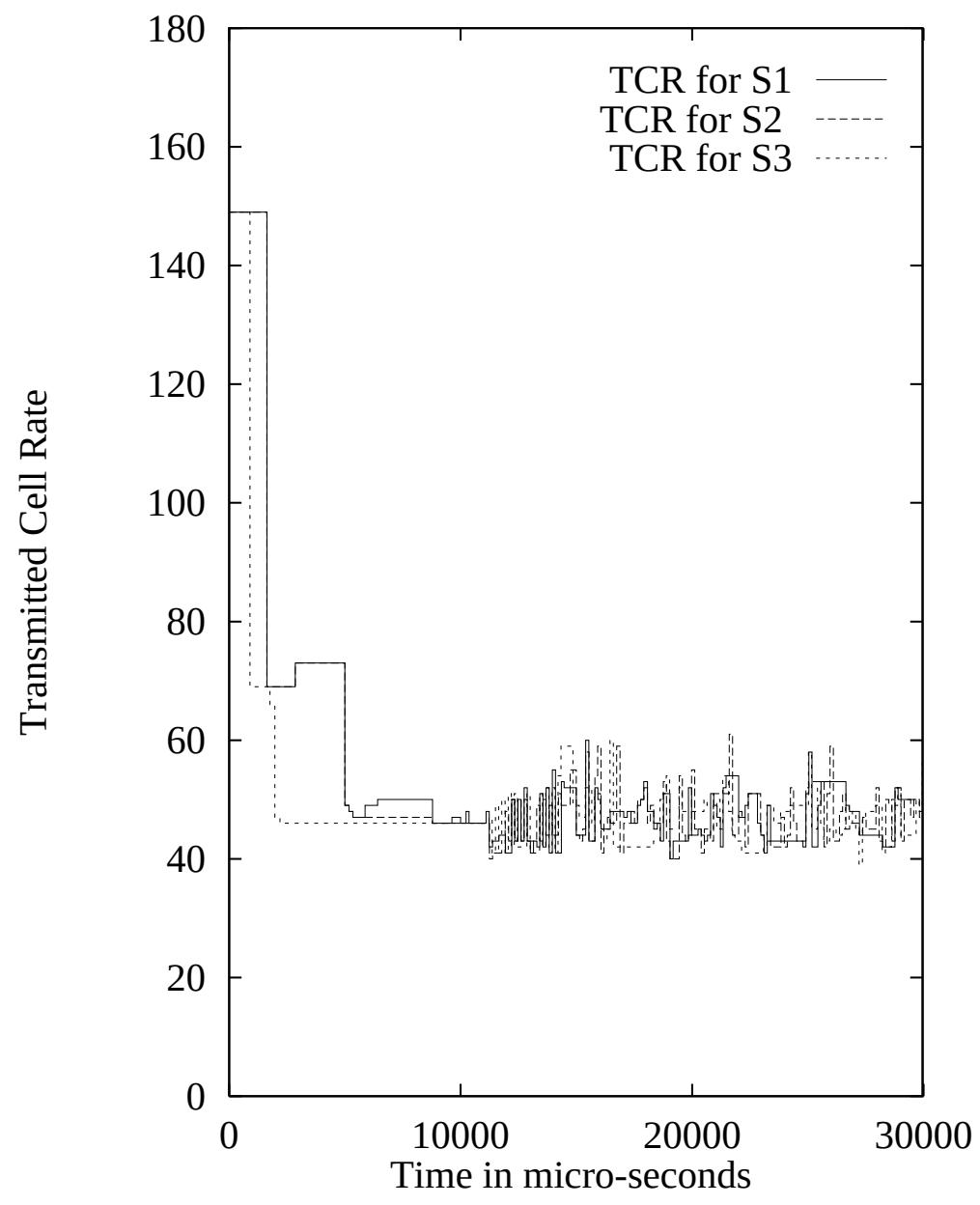

(a) Transmitted Cell Rates

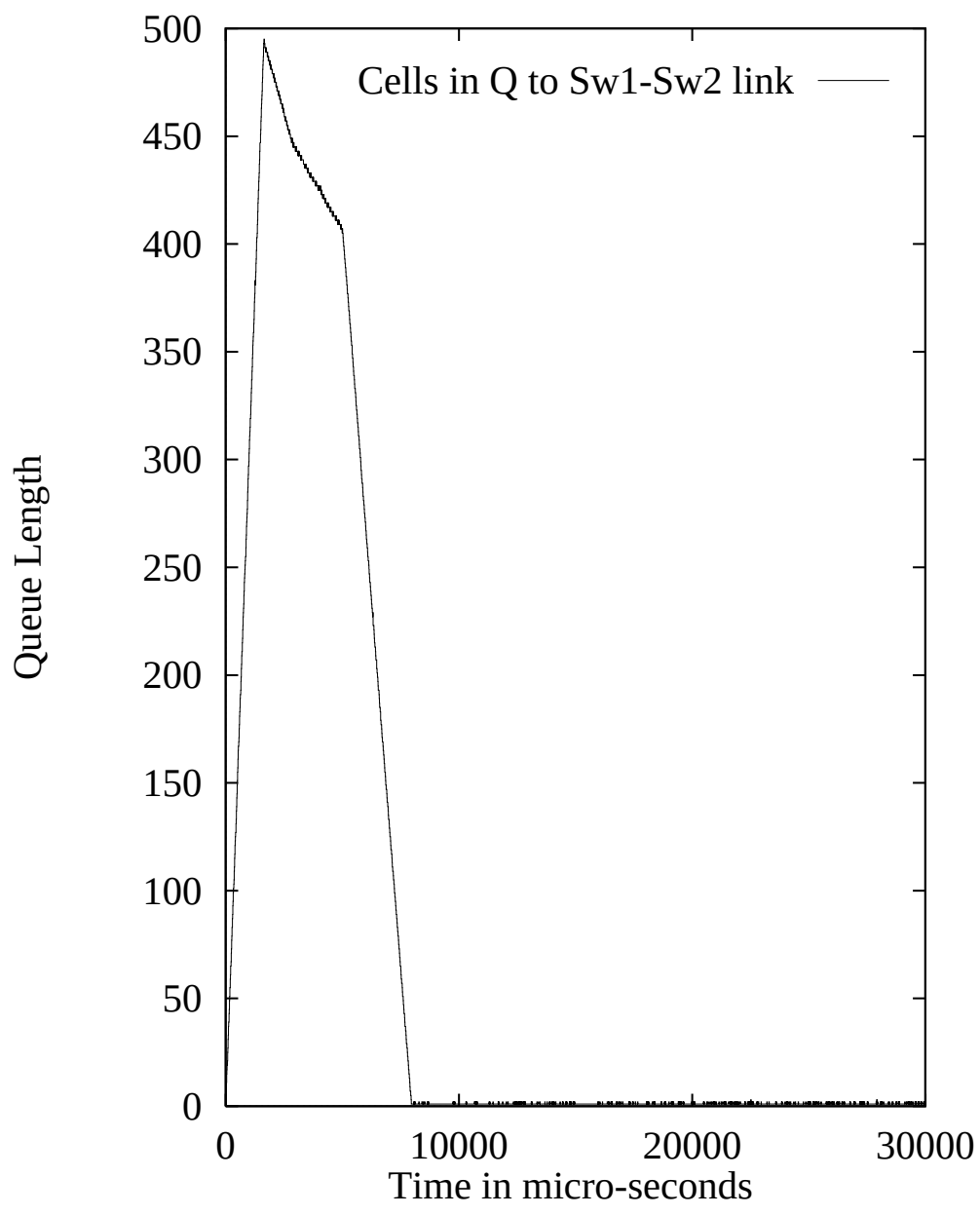

(b) Queue Lengths

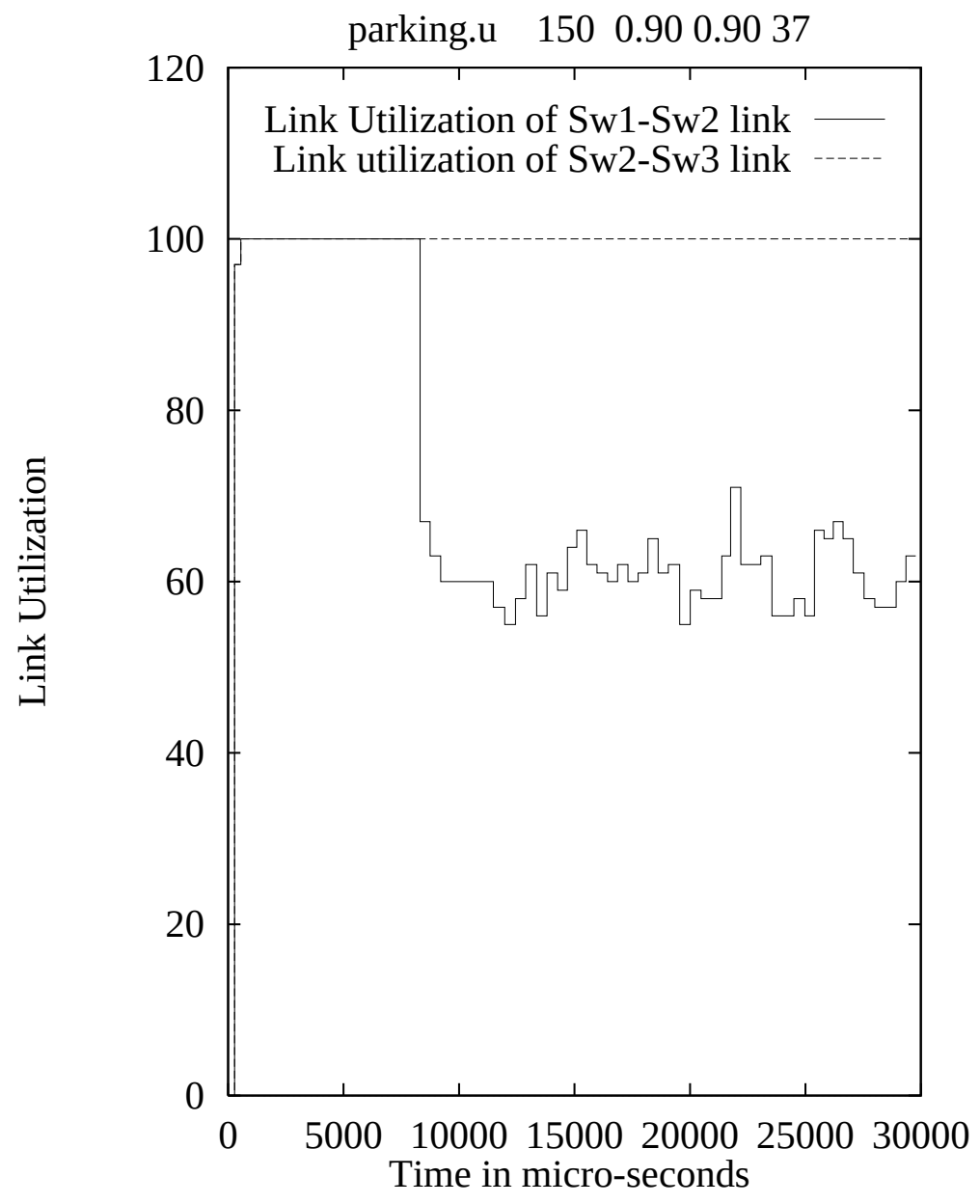

(c) Link Utilization

Figure 7: Simulation results for the parking lot configuration

# 5 Summary

We have developed an end-to-end rate-based congestion avoidance scheme for ABR traffic on ATM networks. The scheme uses a new congestion detection technique and an O(1) switch algorithm and achieves the goals of high thoughput, low queues and fair operation with a small set of parameters whose effects are well understood. A sample set of LAN and WAN simulations are presented.

The OSU scheme has helped shape the traffic management specifications for the available bit rate service. A number of features like congestion avoidance and input rate as the congestion metric have been adoped by other switch schemes.

The scheme, however is not directly compatible with the current ATM Forum Traffic Management standards. The OSU scheme sends control cells every $T$ interval. The is called a time-based approach. The ATM Forum has decided to support a count-based alternative where send control cells after every $n$ data cells. The AI field is not required with the count-based approach. The ATM Forum also uses only one rate which corresponds to TCR in our control cell format. The OCR field is not used.

Though it cannot be used directly, its features can be made compatible. We have developed such a scheme called ERICA which is also mentioned in the ATM Traffic Management 4.0 standards (yet to be published) and will be the subject of our future publications.

[3] Throughout this section, AF-TM refers to ATM Forum Traffic Management sub-working group contributions.

[4] All our papers and ATM Forum contributions are available through http://www.cis.ohio-state.edu/~jain/

# A Detailed Pseudocode

We describe the pseudo code as sets of actions taken when certain events happen at the source adaptor and at the switch.

## A.1 The Source Algorithm

1. Initialization:

   TCR ←Initial Cell Rate;

   Averaging_Interval ←Some initial value;

2. A data cell or cell burst is received from the host.

   Enqueue the cell(s) in the output queue.

3. The inter-cell transmission timer expires.

   IF Output_Queue NOT Empty THEN dequeue the first cell and transmit;

   Increment Transmitted_Cell_Count;

   Restart Inter_Cell_Transmission_Timer;

4. The averaging interval timer expires.

   Offered_Cell_Rate ←Transmitted_Cell_Count/Averaging_Interval;

   Transmitted_Cell_Count ←0;

   Create a control cell;

   OCR_In_Cell ←Offered_Cell_Rate ;

   TCR_In_Cell ←max{TCR, OCR} ;

Load_Adjustment_Factor ←0;

Transmit the control cell;

Restart Averaging_Interval_Timer;

5. A control cell returned from the destination is received.

    New_TCR ←TCR_In_Cell/Load_Adjustment_Factor_In_Cell;

    IF Load_Adjustment_Factor_In_Cell ≥ 1

        THEN IF New_TCR < TCR THEN TCR ←New_TCR ;

        ELSE IF Load_Adjustment_Factor_In_Cell < 1

            THEN IF New_TCR > TCR THEN TCR ←New_TCR ;

    Inter_Cell_Transmission_Time ←1/TCR;

    Averaging_Interval ←Averaging_Interval_In_Cell;

## A.2 The Switch Algorithm

1. Initialization:

    Target_Cell_Rate ←Link_Bandwidth × Target_Utilization / Cell_Size ;

    Target_Cell_Count ←Target_Cell_Rate × Averaging_Interval;

    Received_Cell_Count ←0;

    Clear VC_Seen_Bit for all VCs;

    Upper_Load_Bound ←1 + Half_Width_Of_TUB;

    Lower_Load_Bound ←1 - Half_Width_Of_TUB;

2. A data cell is received.

    Increment Received_Cell_Count;

    Mark VC_Seen_Bit for the VC in the Cell;

3. The averaging interval timer expires.

    Num_Active_VCs ←max{$\sum$ VC_Seen_Bit, 1};

    Fair_Share_Rate ←Target_Cell_Rate/Num_Active_VCs;

    Load_Level ←Received_Cell_Count/Target_Cell_Count;

Reset all VC_Seen_Bits;

Received_Cell_Count ←0;

Restart Averaging_Interval_Timer;

4. A control cell is received.

IF (Load_Level ≥ Lower_Load_Bound) and (Load_Level ≤ Upper_Load_Bound)

THEN IF OCR_In_CELL > Fair_Share_Rate

THEN Load_Adjustment_Decision ←Load_Level/Lower_Load_Bound

ELSE Load_Adjustment_Decision ←Load_Level/Upper_Load_Bound

ELSE Load_Adjustment_Decision ←Load_Level;

IF (Load_Adjustment_Decision > Load_Adjustment_Factor_In_Cell)

THEN Load_Adjustment_Factor_In_Cell ←Load_Adjustment_Decision;